\documentclass[12pt, bibliography=totoc, toc=bibnumbered]{article}
\usepackage{graphicx}

\newcommand\pubnumber{}
\newcommand\pubdate{\today}

\def\institute{DESY\\
Notkestrasse 85, Hamburg, GERMANY}

\def\Title#1{\begin{center} {\Large #1 } \end{center}}
\def\Author#1{\begin{center}{ \sc {#1} on behalf of the CMS collaboration} \end{center}}
\def\Address#1{\begin{center}{ \it #1} \end{center}}

\newcommand\pubblock{\rightline{\begin{tabular}{l} \pubnumber\\
         \pubdate  \end{tabular}}}
\newenvironment{Abstract}{\begin{quotation}  }{\end{quotation}}
\newenvironment{Presented}{\begin{quotation} \begin{center} 
             PRESENTED AT\end{center}\bigskip 
      \begin{center}\begin{large}}{\end{large}\end{center} \end{quotation}}

\def\beq{\begin{equation}}
\def\eeq#1{\label{#1}\end{equation}}
\def\eeqn{\end{equation}}

\def\beqa{\begin{eqnarray}}
\def\eeqa#1{\label{#1}\end{eqnarray}}
\def\eeqan{\end{eqnarray}}

\let\bar=\overbar

\def\Dslash{\not{\hbox{\kern-4pt $D$}}}
\def\dslash{\not{\hbox{\kern-2pt $\del$}}}

\def\msb{{\bar{\ssstyle M \kern -1pt S}}}

\begin{document}
\begin{titlepage}
\pubblock

\vfill
\Title{First measurement of the top quark pair production
cross section at $\sqrt{s} = 13.6 \, \mathrm{TeV}$ at the CMS experiment}
\vfill
\Author{ Laurids Jeppe}
\Address{\institute}
\vfill
\begin{Abstract}
      We present the first measurement of the top quark pair production cross section at the new LHC center-of-mass
      energy of $\sqrt{s} = 13.6 \, \mathrm{TeV}$, using $1.20 \, \mathrm{fb}^{-1}$ of data recorded at the CMS detector. We use a new method combining dilepton and lepton+jets decay channels, constraining several experimental uncertainties in situ. A cross section of $887^{+43}_{-41}(\mathrm{stat+syst}) \pm 53 (\mathrm{lumi}) \, \mathrm{pb}$ is measured, in agreement with the standard model. This result constitutes a first validation of the new data taken by CMS in LHC Run 3.
\end{Abstract}
\vfill
\begin{Presented}
$15^\mathrm{th}$ International Workshop on Top Quark Physics\\
Durham, UK, 4--9 September, 2022
\end{Presented}
\vfill
\end{titlepage}
\def\thefootnote{\fnsymbol{footnote}}
\setcounter{footnote}{0}
\newcommand{\ttbar}{\mathrm{t\bar{t}}}
\newcommand{\Pp}{\mathrm{p}}
\newcommand{\TeV}{\mathrm{TeV}}

\section{Introduction}

Recently, the Large Hadron Collider (LHC) at CERN has reached a new, unprecedented center-of-mass energy of $\sqrt{s} = 13.6 \, \mathrm{TeV}$, starting LHC Run 3. This presents the opportunity to re-measure relevant physical quantities at the new energy frontier, thereby checking the predictions of the standard model (SM).

Here, we present the first ever measurement of the top quark pair ($\ttbar$) production cross section in proton-proton collisions \cite{CMS:TOP-22-012}. Compared to the previous center-of-mass energy of $\sqrt{s} = 13 \, \mathrm{TeV}$, the cross section is expected to rise by about 10\% to $921\,^{+18}_{-16}\,(\mathrm{scale})\pm4\,(\mathrm{PDF+}\alpha_s) \, \mathrm{pb}$. The measurement was performed using the CMS detector \cite{CMS:Detector-2008}, which underwent several upgrades inbetween the end of Run 2 and the beginning of Run 3. Since the detection of top quarks covers a wide range of physics objects reconstructed in CMS, this measurement can also be seen as a first test of the CMS calibration and performance on the new data.

\section{Measurement strategy}

Data corresponding to an integrated luminosity of $1.20 \pm 0.07 \, \mathrm{fb}^{-1}$ are analyzed. Events are selected with two charged leptons (electrons and muons) of opposite charge (dilepton channel), and with a single lepton (lepton+jets channel). In order to improve the separation between signal and background events, the presence of hadronic jets, as well as jets identified as originating from b quarks (b-tagged jets), is required. The inclusion of both decay channels in the analysis allows for constraining lepton selection efficiencies, making the measurement independent from external input on calibrations. Events are categorized by the number and flavours of the leptons, number of jets and b-tagged jets.

\begin{figure}[!tb]
      \centering
      \includegraphics[width=0.49\textwidth]{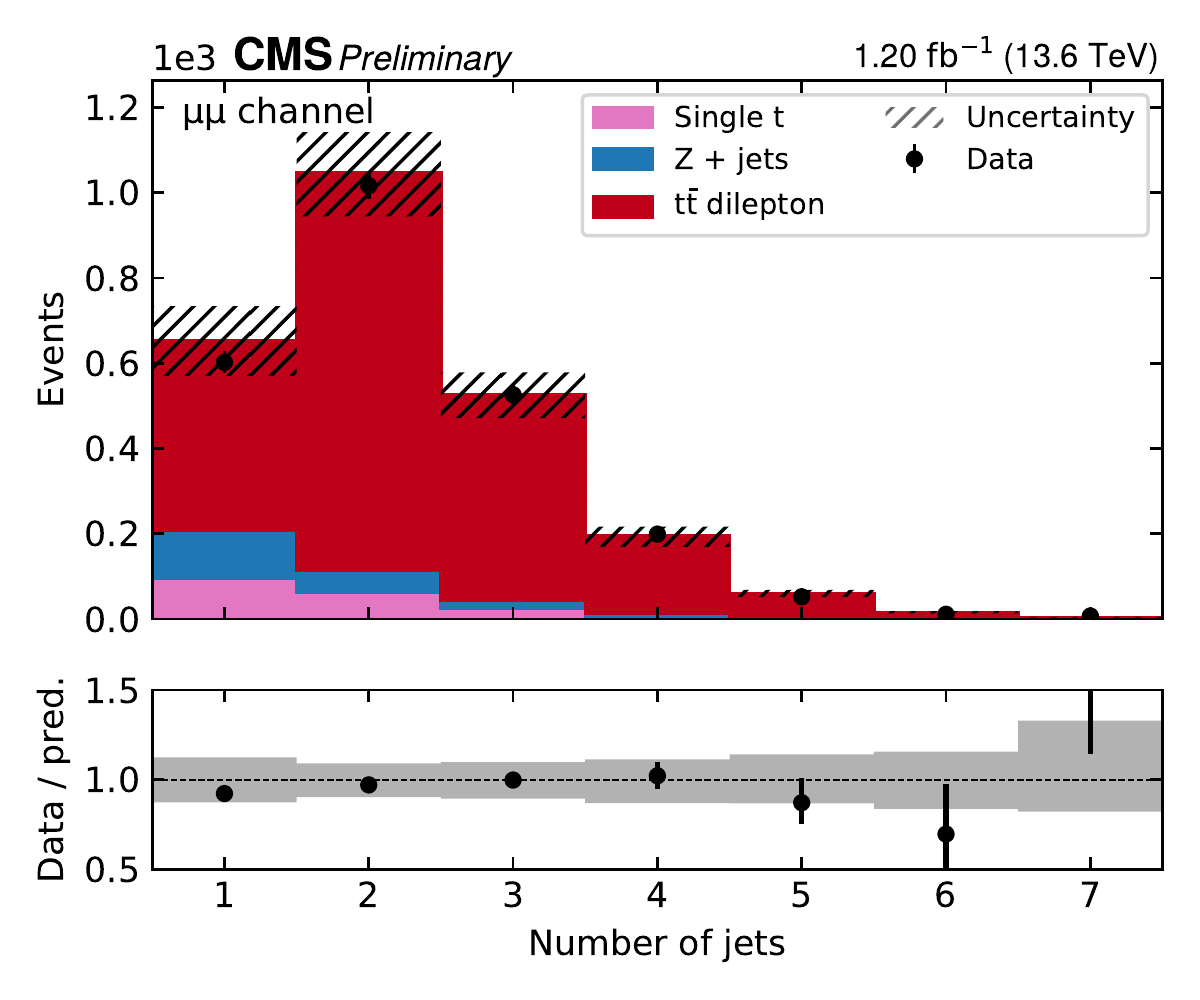}
      \hfill
      \includegraphics[width=0.49\textwidth]{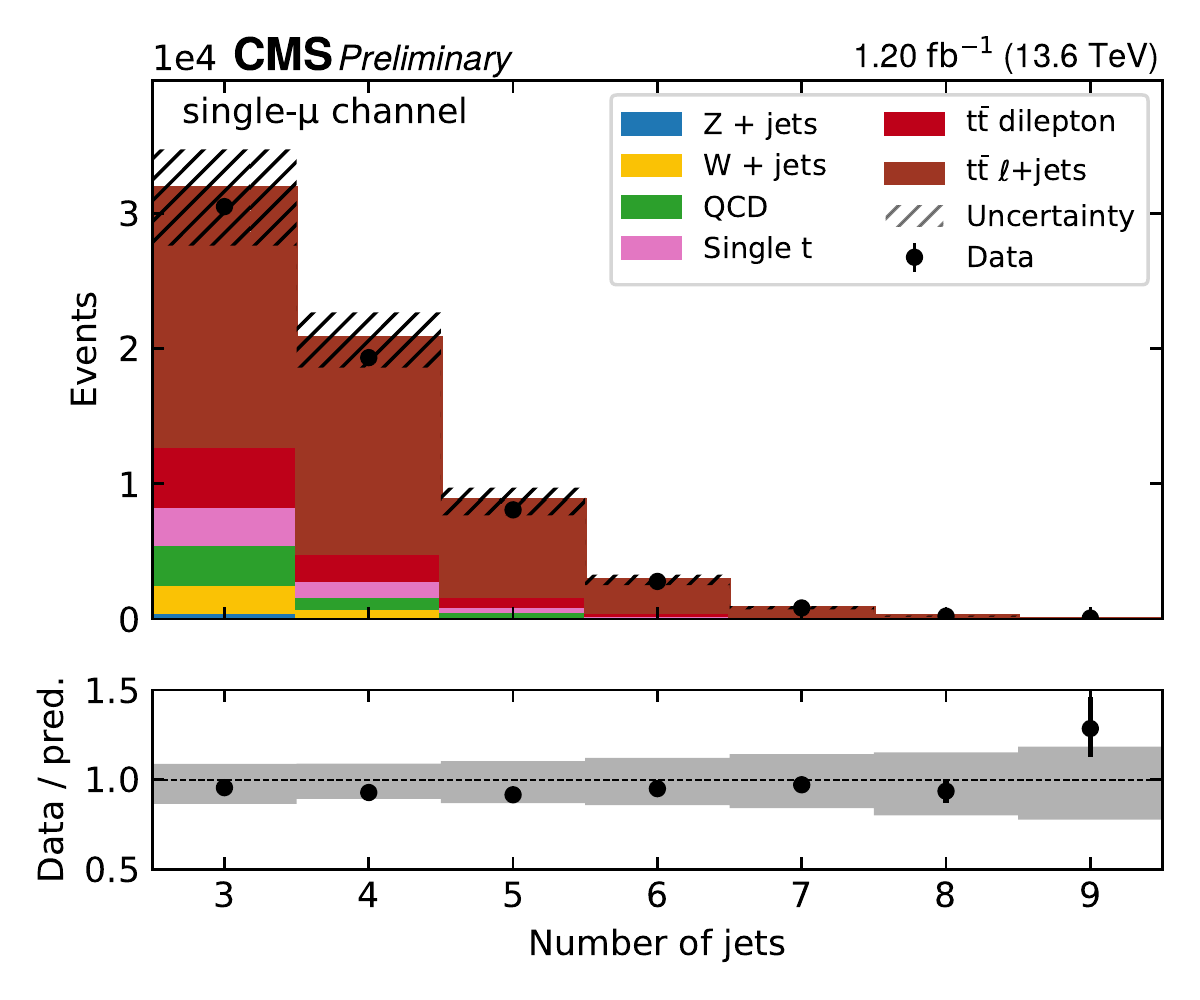}
      \caption{%
      The number of jets for data and simulation in the dimuon (left) and the muon+jets channel (right) \cite{CMS:TOP-22-012}.
      }
      \label{fig:control_plots}
\end{figure}

The $\ttbar$ signal as well as background contributions are estimated using Monte Carlo simulation, with the exception of background originating from non-prompt leptons in QCD events, which is estimated using a data-driven method. The hadronic decay $\mathrm{t} \rightarrow \mathrm{b W} \rightarrow \mathrm{b q q'}$ in the lepton+jets channels is utilized to derive a coarse calibration of the jet energy by constructing a dijet mass distribution and using it to correct the simulation to data. Figure \ref{fig:control_plots} shows, as an example, the agreement between data and simulation in the dimuon and muon+jets channels.

\begin{figure}[!tbp]
      \centering
      \includegraphics[width=0.8\textwidth]{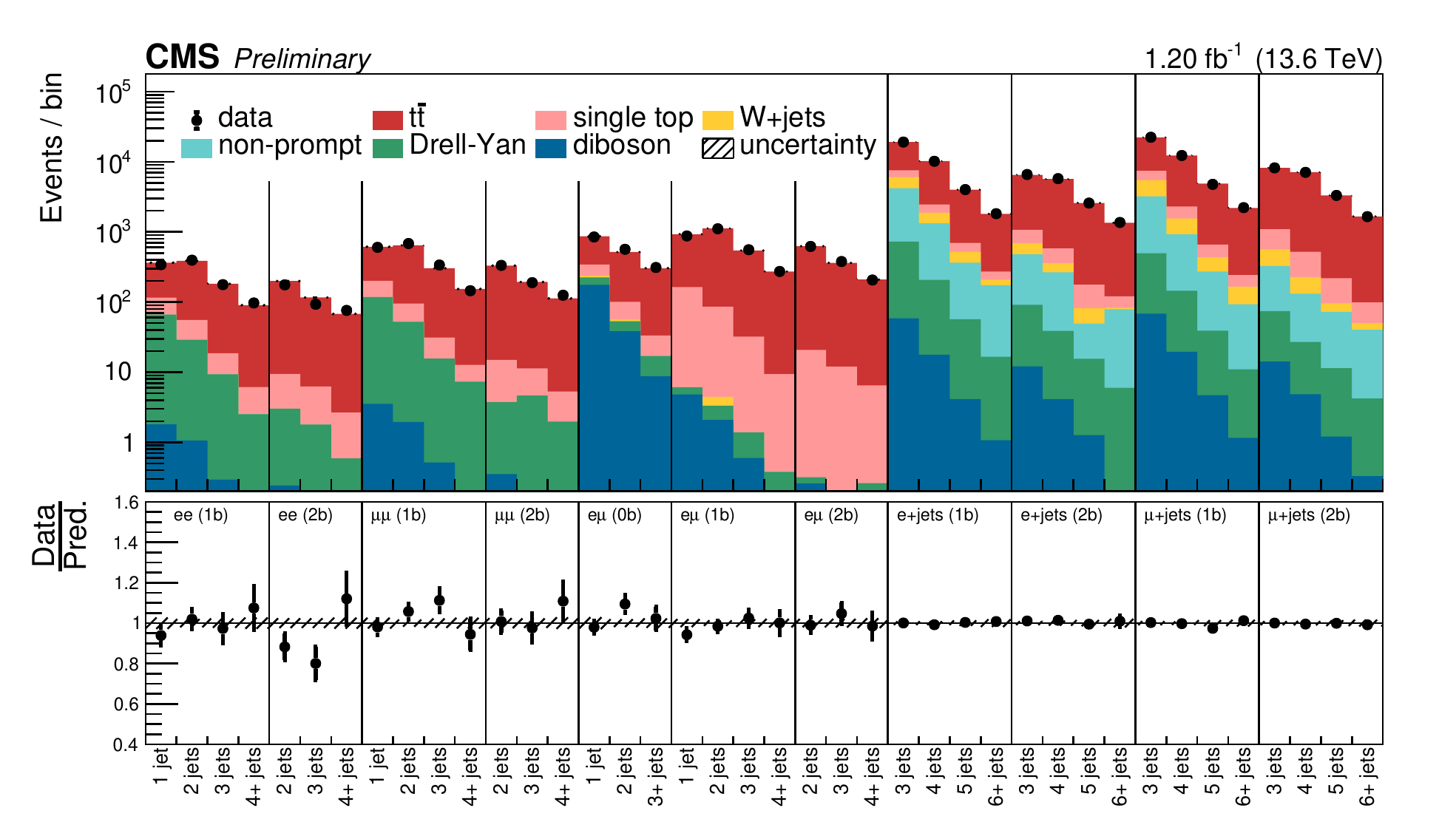}
      \caption{
            Resulting post-fit agreement and uncertainties in the fit binning after performing the fit \cite{CMS:TOP-22-012}.
      }
      \label{fig:postfit}
\end{figure}

To carry out the measurement of the $\ttbar$ cross section, events are further binned by the number of jets. A maximum likelihood fit over all event categories is performed, and systematic uncertainties are included in the fit as nuisance parameters. This way, both lepton selection and b-tagging efficiencies are determined directly in-situ together with the cross section. The uncertainty on the luminosity is not directly considered in the fit, and instead treated externally. Figure \ref{fig:postfit} shows the post-fit agreement of all categories in the final fit binning.

\section{Result}

\begin{figure}[!ht]
      \centering
      \includegraphics[width=0.8\textwidth]{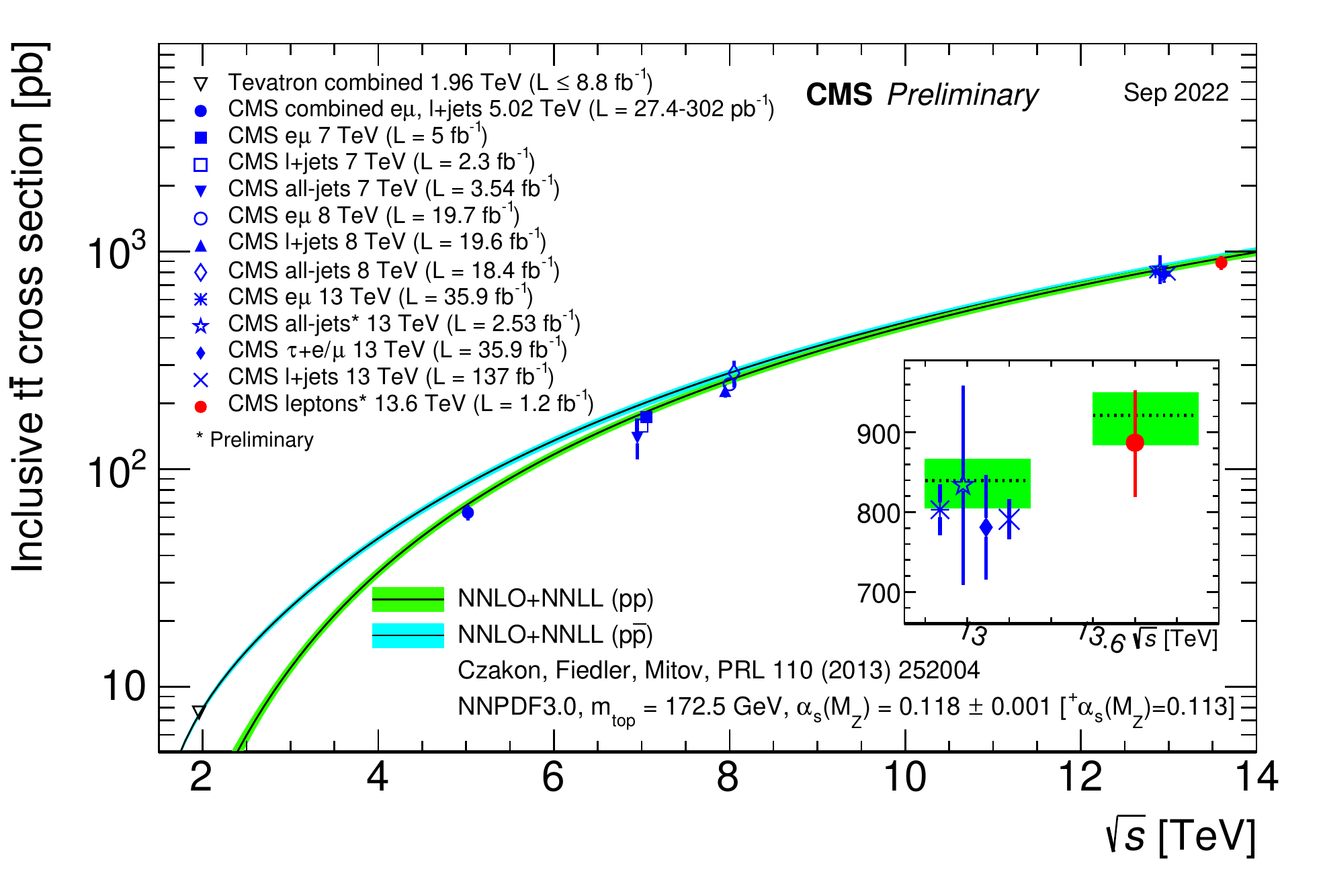}
      \caption{%
          The $\ttbar$ cross section as a function of $\sqrt{s}$, as obtained in previous measurements by the CMS experiment in pp collisions (blue markers) and at the Tevatron in $\mathrm{p\bar{p}}$ collisions (empty black inverted triangle).
          The red bullet shows the result from this analysis at $\sqrt{s} = 13.6 \, \mathrm{TeV}$ \cite{CMS:TOP-22-012}.
      }
      \label{fig:xsec}
\end{figure}

From the likelihood fit, a cross section of $887^{+43}_{-41}(\mathrm{stat+syst}) \pm 53 (\mathrm{lumi}) \, \mathrm{pb}$ is measured. A comparison to previous measurements from CMS at lower center-of-mass energies, as well as the SM prediction, can be seen in Fig. \ref{fig:xsec}. The statistical and systematic error encompass the full uncertainty of the likelihood fit, while the error on the luminosity is estimated from emittance scans \cite{CMS:DP-2018-011}. The result is in good agreement with the SM, and shows excellent performance from the first CMS data in Run 3.

\bibliography{literature}
\bibliographystyle{unsrt}

\end{document}